\documentclass[a4paper,conference]{IEEEtran}
\IEEEoverridecommandlockouts
\usepackage{cite}
\usepackage{amsmath,amssymb,amsfonts}
\usepackage{algorithmic}
\usepackage{graphicx}
\usepackage{textcomp}
\usepackage{enumerate}
\usepackage{xcolor}
\usepackage{bm}
\usepackage{multirow}
\usepackage{url}
\def\BibTeX{{\rm B\kern-.05em{\sc i\kern-.025em b}\kern-.08em
    T\kern-.1667em\lower.7ex\hbox{E}\kern-.125emX}}
\begin{document}

\title{Multi-color balancing for correctly adjusting\\ the intensity of target colors
\thanks{}
}

\author{\IEEEauthorblockN{1\textsuperscript{st} Teruaki Akazawa}
\IEEEauthorblockA{\textit{Tokyo Metropolitan University}\\
Tokyo, Japan \\
akazawa-teruaki@ed.tmu.ac.jp}
\and
\IEEEauthorblockN{2\textsuperscript{nd} Yuma Kinoshita}
\IEEEauthorblockA{\textit{Tokyo Metropolitan University}\\
Tokyo, Japan \\
ykinoshita@tmu.ac.jp}
\and
\IEEEauthorblockN{3\textsuperscript{rd} Hitoshi Kiya}
\IEEEauthorblockA{\textit{Tokyo Metropolitan University}\\
Tokyo, Japan \\
kiya@tmu.ac.jp}
}

\maketitle

\begin{abstract}
In this paper, we propose a novel multi-color balance method for reducing color distortions caused by lighting effects.
The proposed method allows us to adjust three target-colors chosen by a user in an input image so that each target color is the same as the corresponding destination (benchmark) one.
In contrast, white balancing is a typical technique for reducing the color distortions, however, they cannot remove lighting effects on colors other than white.
In an experiment, the proposed method is demonstrated to be able to remove lighting effects on selected three colors, and is compared with existing white balance adjustments.
\end{abstract}
\section{Introduction}
In human perception, even if illumination conditions change, the color appearance of objects in a scene does not change\cite{color_managemanagement_sys}.
White balance adjustment is a technique for reproducing this ability as a computer vision task\cite{AWB_goes_wrong,Deep_WB_Editing,brucelindbloom}.
It is generally performed to remove the effects of illumination conditions, i.e., lighting effects.
To apply white-balancing to images, first, we need to estimate a white region with remaining lighting effects, called ``source white point.''
There are many studies on estimating source white points\cite{Max_RGB,Grey_World_Theory,Grey_Edge,Cheng_PCA,APAP_bias_illumination_estimation,AWB_goes_wrong,Deep_WB_Editing}.
Next to the estimation, a color transform matrix, which maps a source white point into a desired destination white point, are designed, and then the matrix is applied to each pixel in an image.
Once a source white point is correctly estimated and white-balancing is applied to an image, lighting effects on white are perfectly removed.
However, even when a white balance adjustment is applied, there are still some lighting effects on other colors.
This is because colors other than white are not considered in designing the matrix\cite{ChengsBeyondWhite}.
For this reason, in some computer vision tasks such as detecting objects with specific colors, the specific colors still suffer from lighting effects\cite{Fast_Soft_Segmentation_2020_CVPR,seo-san-Hue_Correction_Scheme_2020,kinoshita-san-APSIPA-HUE,kinoshita-san-Hue,kinoshita-san-Semantic_2019,Color_Constancy_effect_on_DNN_2019_ICCV,kinoshita-san-Automatic_Exposure_Compensation_2018}.

Accordingly, in this paper, we propose a target-color correction method based on a novel multi-color balance adjustment.
In the proposed multi-color balancing, a color transform matrix is designed from three target colors.
By applying the proposed method to an input image, each target color in the image is perfectly adjusted to the corresponding desired color.

In an experiment, the effectiveness of the proposed method is shown, compared with state-of-the-art white-balance adjustments.
\section{Related work}
Pixel values captured by an RGB digital camera are given by using three elements: spectra of illumination, spectral reflectance of objects and camera spectral sensitivity\cite{Hirai}.
For this reason, illumination changes affect the color of captured images. 
Hence, white-balancing has generally been applied to images to remove the lighting effects, so far\cite{brucelindbloom}.
\subsection{White balance adjustment}
Let ${\bm{P}_{\rm{XYZ}}=(X_{\rm{P}},Y_{\rm{P}},Z_{\rm{P}})^\top}$ and ${\bm{P}'_{\rm{XYZ}}=(X'_{\rm{P}},Y'_{\rm{P}},Z'_{\rm{P}})^\top}$ be a pixel value of input image ${I}_{\rm{XYZ}}$ in the XYZ color space\cite{CIE_XYZ_color_space} and that of the corresponding white-balanced image ${I}'_{\rm{XYZ}}$, respectively.
A white balance adjustment is performed by the following equation\cite{brucelindbloom}:
\begin{equation}
\label{eqn:chromadaptWB}
\bm{P}'_{\rm{XYZ}} = \bf{{M}_{\rm{WB}}} \bm{{P}}_{\rm{XYZ}}
.
\end{equation}
Matrix $\bf{M}_{\rm{WB}}$ in (\ref{eqn:chromadaptWB}) is given as
\begin{equation} 
\label{eqn:Mwb}
    \bf{M}_{\rm{WB}} = {\bf{M}_{\rm{A}}}^{\rm{-1}}
    \left(
    \begin{array}{cccc}
    {\rho_{\rm{D}}}/{\rho_{\rm{S}}} & 0 & 0 \\
   0 & {\gamma_{\rm{D}}}/{\gamma_{\rm{S}}} & 0 \\
   0 & 0 & {\beta_{\rm{D}}}/{\beta_{\rm{S}}} \\
    \end{array}
    \right)
    \bf{M}_{\rm{A}}
,
\end{equation}
where $(\rho_{\rm{S}},\gamma_{\rm{S}},\beta_{\rm{S}})^\top$ and $(\rho_{\rm{D}},\gamma_{\rm{D}},\beta_{\rm{D}})^\top$ are calculated from a source white point $(X_{\rm{S}},Y_{\rm{S}},Z_{\rm{S}})^\top$ of image ${I}_{\rm{XYZ}}$ and a desired destination (benchmark) white point $(X_{\rm{D}},Y_{\rm{D}},Z_{\rm{D}})^\top$ as
\begin{equation}
\label{eqn:color-xfer}
\left(
\begin{array}{cccc}
   {\rho_{\rm{S}}} \\
   {\gamma_{\rm{S}}} \\
   {\beta_{\rm{S}}} \\
  \end{array} 
  \right)
  = \bf{M}_{\rm{A}} \left(
  \begin{array}{cccc}
   {X_{\rm{S}}} \\
   {Y_{\rm{S}}} \\
   {Z_{\rm{S}}} \\
  \end{array}
    \right) \: , \:
 \left(
\begin{array}{cccc}
   {\rho_{\rm{D}}} \\
   {\gamma_{\rm{D}}} \\
   {\beta_{\rm{D}}} \\
  \end{array} 
  \right)
  = \bf{M}_{\rm{A}} \left(
  \begin{array}{cccc}
   {X_{\rm{D}}} \\
   {Y_{\rm{D}}} \\
   {Z_{\rm{D}}} \\
  \end{array}
    \right) 
\end{equation}
Note that some automatic algorithms are available for estimating source white point $(X_{\rm{S}},Y_{\rm{S}},Z_{\rm{S}})^\top$\cite{Max_RGB,Grey_World_Theory,Grey_Edge,Cheng_PCA,APAP_bias_illumination_estimation,AWB_goes_wrong,Deep_WB_Editing}.
Matrix $\bf{M}_{\rm{A}}$ with a size of 3$\times$3 is decided in accordance with an assumed chromatic adaption model.
When white-balancing is directly carried out in the XYZ color space, $\bf{M}_{\rm{A}}$ will be the 3$\times$3-identity matrix.
We call this white-balancing ``white-balancing with XYZ scaling'' in this paper.
In contrast, in Bradford’s\cite{BradfordOriginal} or von Kries’s\cite{vonKriesOriginal} cone response model, $\bf{M}_{\rm{A}}$ does not correspond to the 3$\times$3-identity matrix for high-quality white-balancing.
For example, under the use of Bradford's model, $\bf{M}_{\rm{A}}$ is given as
\begin{equation}  
\label{eqn:Ma} 
  \bf{M}_{\rm{A}} = \left(
  \begin{array}{cccc}
   0.8951 & 0.2664 & -0.1614 \\
   -0.7502 & 1.7135 & 0.0367 \\
   0.0389 & -0.0685 & 1.0296 \\
  \end{array}
    \right)
.
\end{equation}
\subsection{Problem with white balancing}
By using white-balancing, lighting effects on white color can be perfectly removed if the white point of an input image can be estimated correctly\cite{brucelindbloom}.
However, lighting effects on other colors cannot be corrected because colors except white are not considered in the calculation of $\bf{M}_{\rm{WB}}$ (see (\ref{eqn:Mwb})).
Accordingly, in this paper, we propose a novel multi-color balance method that enables us to perfectly remove lighting effects on three target-colors.
\section{Proposed multi-color balancing}
\subsection{Overview}
\begin{figure}[t]
\begin{center}
\includegraphics[keepaspectratio, scale=0.295]{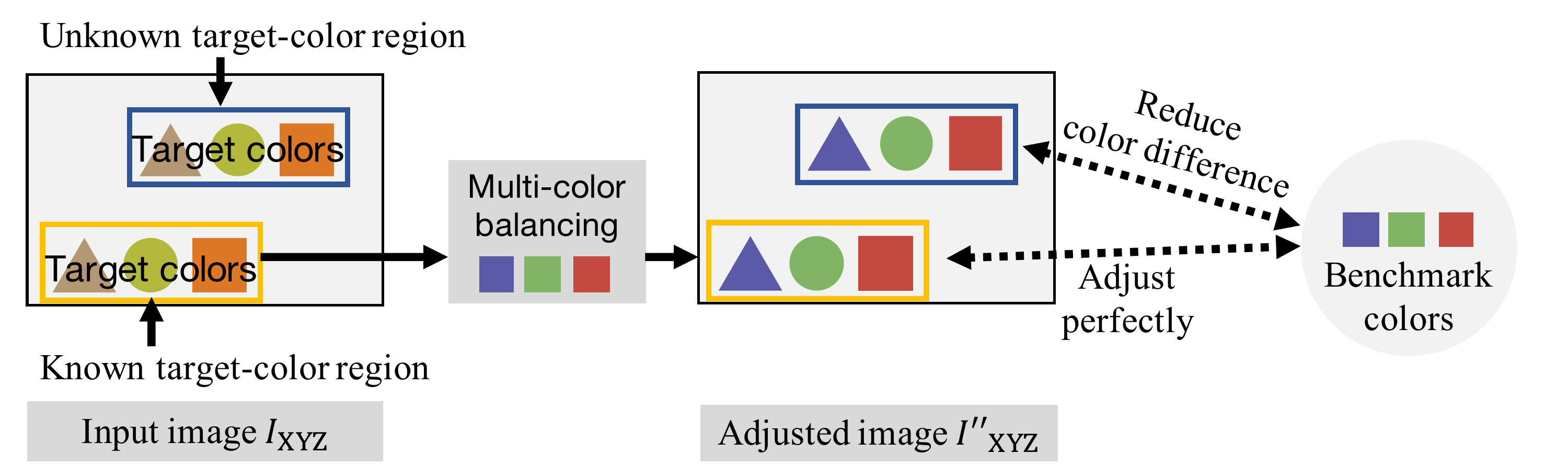}
\end{center}
\caption{Overview of proposed method.}
\label{fig:overview}
\end{figure}
As shown in Fig. \ref{fig:overview}, we assume that objects having target colors are in an input image (${I}_{\rm{XYZ}}$), and their locations are known. 
The locations are called ``known target-color region.''
Besides, we suppose that there are other regions in which objects having the same target-colors as ones in the known region are located.
The regions are called ``unknown target-color region.''
By using a color transform matrix calculated from three target-colors in the known target-color region, the proposed multi-color balancing can adjust target colors not only in the known target-color region but also in the unknown regions to desired destination (benchmark) colors.
\begin{figure}[t]
\begin{center}
\includegraphics[keepaspectratio, scale=0.385]{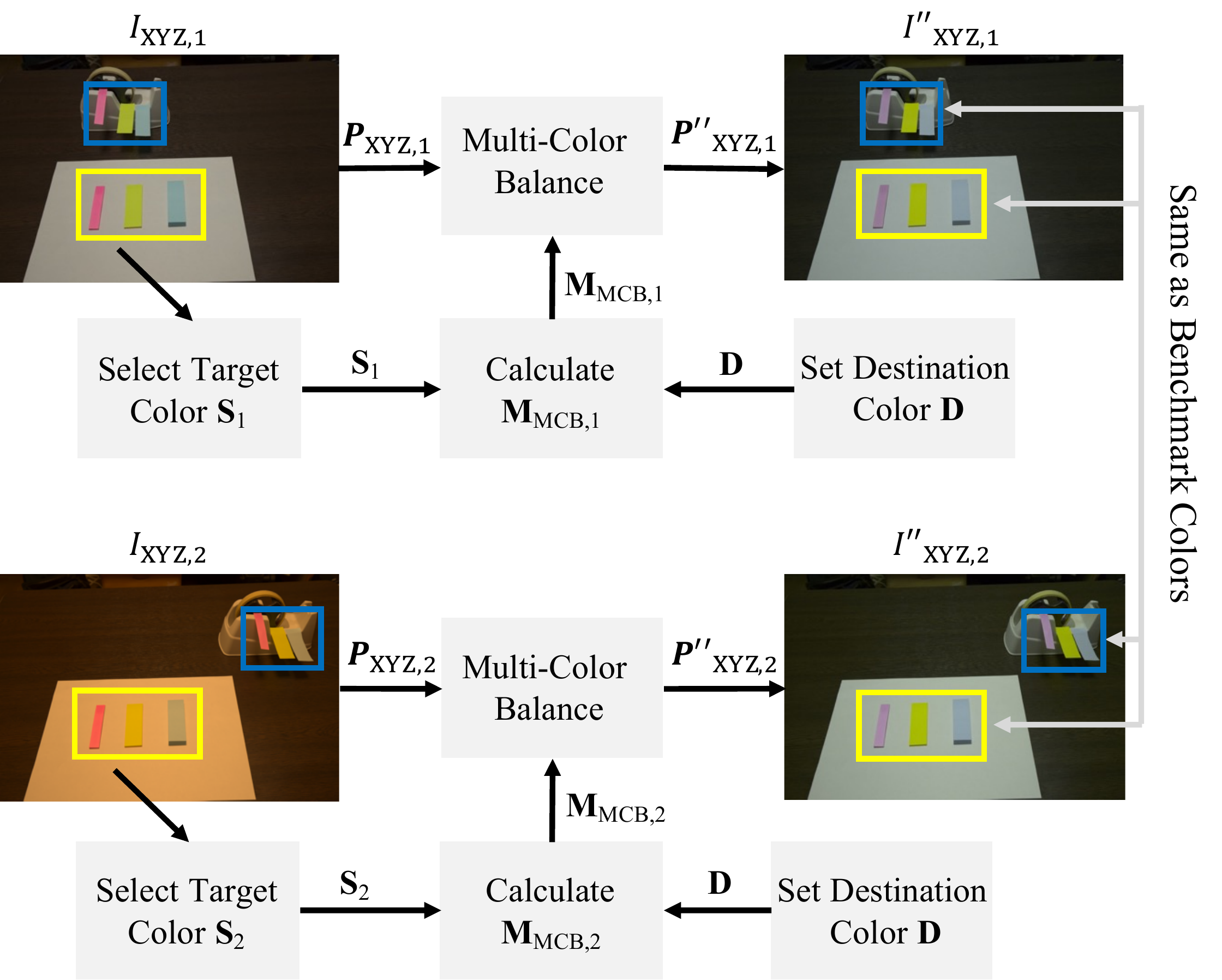}
\end{center}
\caption{Example of applying the proposed method to two images under two different light sources.Yellow rectangle: known target-color region, 
blue rectangle: unknown target-color region.}
\label{fig:method-overview}
\end{figure}
\subsection{Proposed multi-color balancing}
As with white balance, multi-color balanced pixel ${\bm{P}''_{\rm{XYZ}}}$ is given as
\begin{equation}
\label{eqn:chromadapt3}
\bm{P}''_{\rm{XYZ}} = \bf{{M}_{\rm{MCB}}} \bm{{P}}_{\rm{XYZ}}
.
\end{equation}
Let
$(X_{\rm{S1}},Y_{\rm{S1}},Z_{\rm{S1}})^\top$,
$(X_{\rm{S2}},Y_{\rm{S2}},Z_{\rm{S2}})^\top$ and
$(X_{\rm{S3}},Y_{\rm{S3}},Z_{\rm{S3}})^\top$ be three target-colors in the XYZ color space.
Also, let
$(X_{\rm{D1}},Y_{\rm{D1}},Z_{\rm{D1}})^\top$,
$(X_{\rm{D2}},Y_{\rm{D2}},Z_{\rm{D2}})^\top$ and
$(X_{\rm{D3}},Y_{\rm{D3}},Z_{\rm{D3}})^\top$ be corresponding benchmark colors respectively.
Then, $\bf{M}_{\rm{MCB}}$ satisfies
\begin{equation}
\bf{D} = \bf{M}_{\rm{MCB}} \bf{S} 
,
\end{equation}
where
\begin{align}
\label{eqn:S-D_matrix}
\bf{S} = \left(\!\!
  \begin{array}{cccc}
   {X_{\rm{S1}}} & {X_{\rm{S2}}} & {X_{\rm{S3}}} \\
   {Y_{\rm{S1}}} & {Y_{\rm{S2}}} & {Y_{\rm{S3}}} \\
   {Z_{\rm{S1}}} & {Z_{\rm{S2}}} & {Z_{\rm{S3}}} \\
  \end{array}
 \!\!   \right),
\:
\bf{D} = \left(\!\!
  \begin{array}{cccc}
   {X_{\rm{D1}}} & {X_{\rm{D2}}} & {X_{\rm{D3}}} \\
   {Y_{\rm{D1}}} & {Y_{\rm{D2}}} & {Y_{\rm{D3}}} \\
   {Z_{\rm{D1}}} & {Z_{\rm{D2}}} & {Z_{\rm{D3}}} \\
  \end{array}
\!\!    \right)
.
\end{align}
When both $\bf{S}$ and $\bf{D}$ have full-rank, $\bf{M}_{\rm{MCB}}$ is designed by 
\begin{equation}
\label{eqn:Mmcb}
\bf{M}_{\rm{MCB}} = \bf{D} \bf{S}^{\rm{-1}}
.
\end{equation}
When $\bf{M}_{\rm{MCB}}$ in (8) is applied to $I_{\rm{XYZ}}$, a multi-color balanced $I''_{\rm{XYZ}}$ is obtained where target colors in the known target-color region is the same as benchmark colors.
Target colors in the unknown target-color region are reduced to those close to benchmark colors.
%
%
\subsection{Application to two images under two different light sources}
To correct target-colors in two images ($I_{\rm{XYZ,1}}$ and $I_{\rm{XYZ,2}}$) taken under two different light sources, the proposed method is applied to the images as following steps (see Fig. \ref{fig:method-overview}).
\begin{enumerate}[(i)]
\setlength{\itemsep}{5pt}
\item Decide three benchmark colors $(X_{\rm{D1}},Y_{\rm{D1}},Z_{\rm{D1}})^\top$, $(X_{\rm{D2}},Y_{\rm{D2}},Z_{\rm{D2}})^\top$ and $(X_{\rm{D3}},Y_{\rm{D3}},Z_{\rm{D3}})^\top$ in the XYZ color space, and then design  ${\bf{D}}$ by using the benchmark colors, as in (\ref{eqn:S-D_matrix}).
\item Select three target-colors from the known target-color region in $I_{\rm{XYZ,1}}$: $(X_{\rm{S1,1}},Y_{\rm{S1,1}},Z_{\rm{S1,1}})^\top$, $(X_{\rm{S2,1}},Y_{\rm{S2,1}},Z_{\rm{S2,1}})^\top$ and $(X_{\rm{S3,1}},Y_{\rm{S3,1}},Z_{\rm{S3,1}})^\top$, and then define ${\bf{S}}_{1}$ as in (\ref{eqn:S-D_matrix}).\label{item:target-colors}
\item Calculate ${\bf{M}}_{\rm{MCB,1}}$ with  ${\bf{S}}_{1}$ and ${\bf{D}}$, as in (\ref{eqn:Mmcb}).
\item Transform every pixel value ${\bm{P}}_{\rm{XYZ,1}}$ in $I_{\rm{XYZ,1}}$ to obtain multi-color balanced image $I''_{\rm{XYZ,1}}$, as in (\ref{eqn:chromadapt3}).\label{item:end-proposed}
\item Similarly to $I_{\rm{XYZ,1}}$, $I_{\rm{XYZ,2}}$ is transformed in accordance with (\ref{item:target-colors}) -- (\ref{item:end-proposed}).
Note that if the same ${\bf{D}}$ as that of $I_{\rm{XYZ,1}}$ is used, target-colors in $I_{\rm{XYZ,2}}$ are reduced to the same colors as in $I''_{\rm{XYZ,1}}$.
\end{enumerate}
%
%
%
\section{Experiment}
We conducted an experiment to make sure that the proposed method can reduce lighting effects.
\subsection{Experimental conditions}
\label{sec:expprocedure}
\begin{figure}[tb]
\begin{minipage}[b]{0.45\linewidth}
  \centering
  \centerline{\includegraphics[keepaspectratio, scale=0.4]{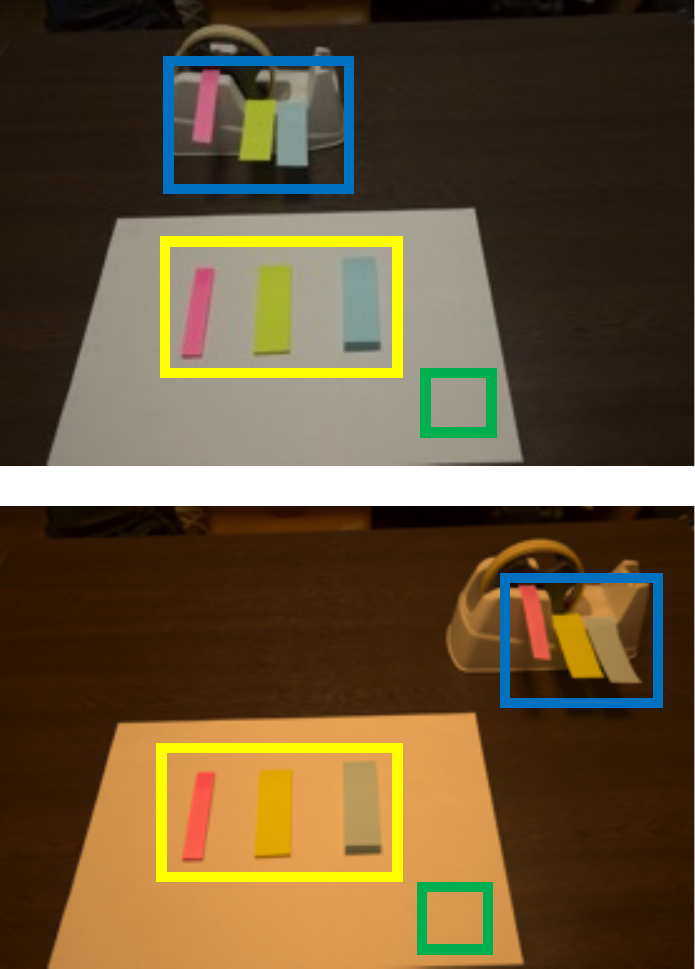}}
  \centerline{(a) Post-it note}\medskip
\end{minipage}
\begin{minipage}[b]{0.5\linewidth}
  \centering
  \centerline{\includegraphics[keepaspectratio, scale=0.4]{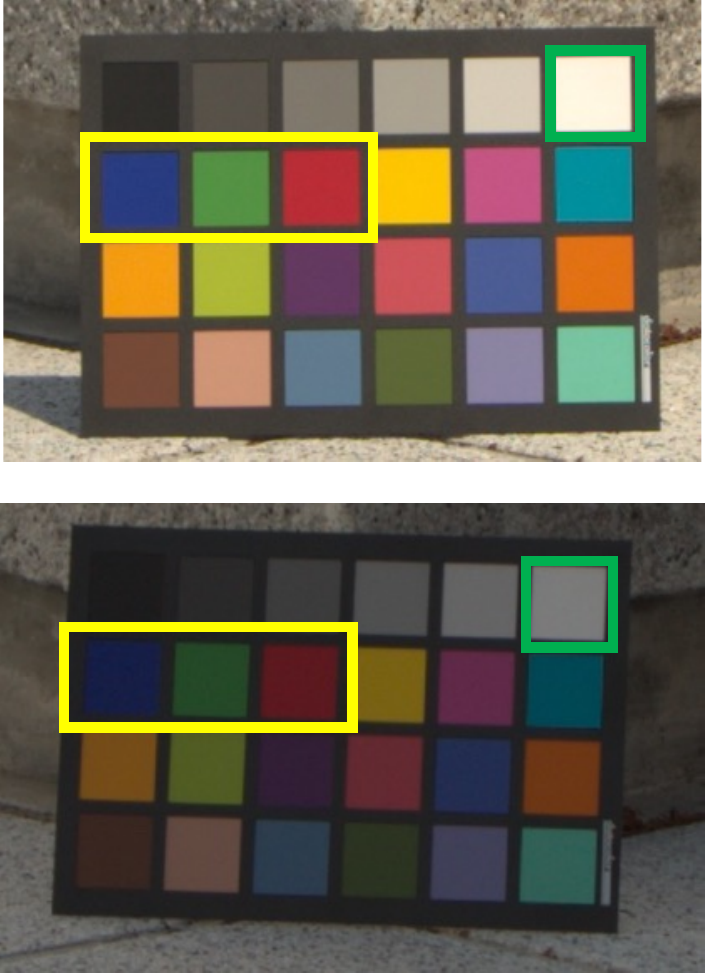}}
  \centerline{(b) Color checker}\medskip
\end{minipage}
%
\caption{Two sets of images used in this experiment.
Yellow rectangle: known target-color region, blue rectangle: unknown target-color region, green rectangle: white region.}
\label{fig:exp_images}
\end{figure}
Two sets of images: ``Post-it note'' and ``Color checker'' were prepared for the experiment (see Fig. \ref{fig:exp_images}).
Each set consists of two images captured in a scene under different lighting conditions, where the two images have the same target colors in a known target-color region, and moreover, have the same objects (post-it notes or a color rendition chart) as shown in Fig. \ref{fig:exp_images}.

The color difference between each benchmark color and the corresponding color in an adjusted image was evaluated by using two metrics as below.
\begin{enumerate}[(i)]
\item Reproduction angular error\cite{ReproductionError}.
\item Hue difference $\Delta H$ of the CIEDE2000 \cite{ISO-CIE_CIEDE2000,CIEDE2000}.
\end{enumerate}

The proposed method was compared with white-balancing with XYZ scaling and white-balancing with Bradford’s cone response model.
For $(X_{\rm{D}},Y_{\rm{D}},Z_{\rm{D}})^\top$ and $(X_{\rm{S}},Y_{\rm{S}},Z_{\rm{S}})^\top$ in (\ref{eqn:color-xfer}), we used the CIE standard illuminant D65\cite{ISO_CIE_standard_illuminant_D65} and the average pixel value of a white region selected manually from each input image (see the green rectangle in Fig. \ref{fig:exp_images}), respectively.
Three target-colors were also selected from the known target-color region, where the average pixel value of each object was used as a target color.
\subsection{Experimental results}
\label{sec:expresults}
%
\begin{figure*}[b]
\begin{center}
\includegraphics[keepaspectratio, scale=0.55]{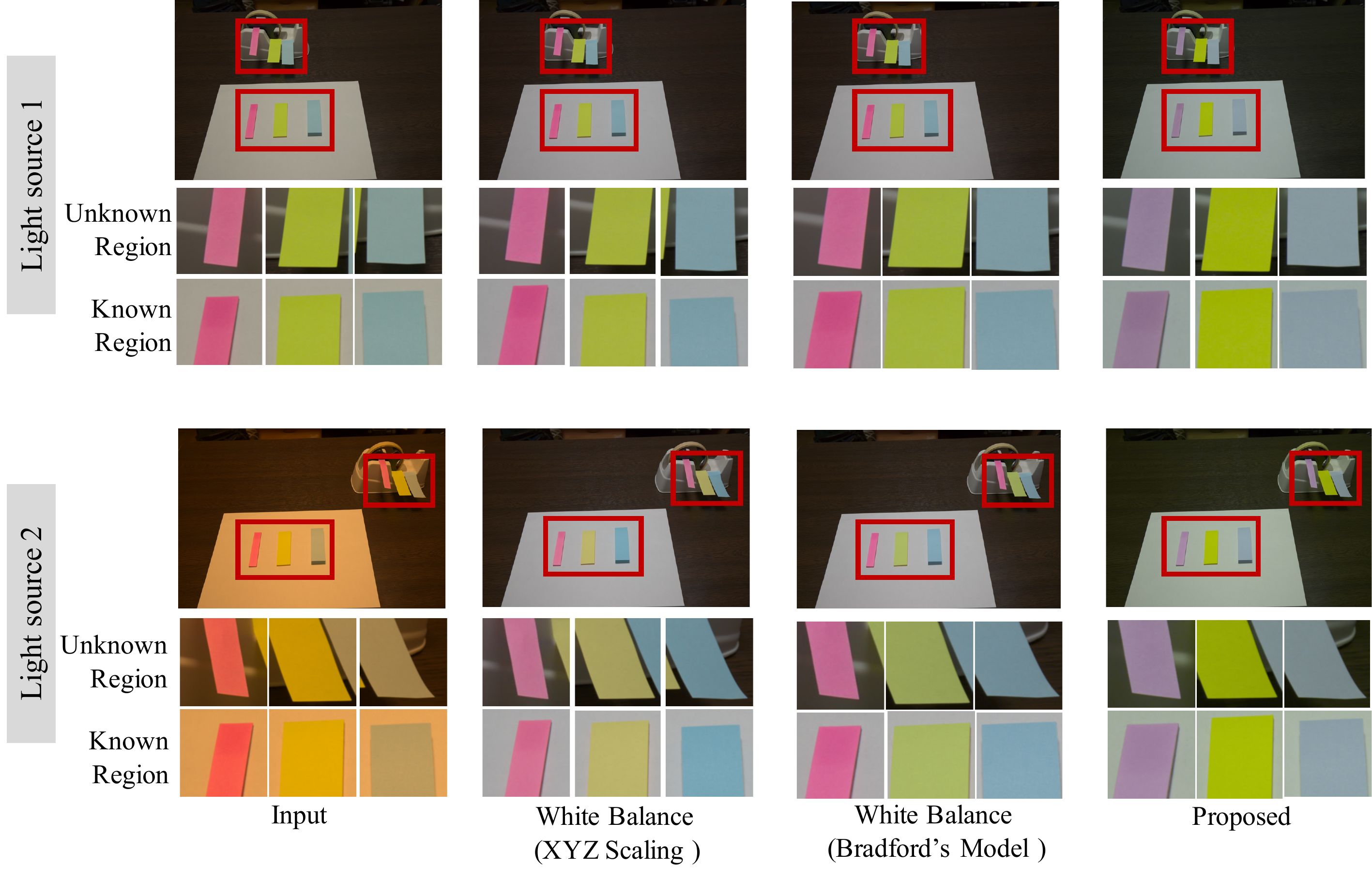}
\end{center}
\caption{Experimental results for ``Post-it note".
Zoom-ins of boxed regions are shown in bottom of each image.}
\label{fig:result-image}
\end{figure*}
%
%
\begin{table}[tb]
\caption{Reproduction angular error (post-it note).}
\label{table:1}
\begin{center}
\scalebox{0.8}{
\begin{tabular}{l|c|ccc|ccc}
\hline
\noalign{\vskip.5mm}
\multirow{2}{*}{Method} &  \multirow{2}{*}{White}  &
\multicolumn{3}{c|}{Known region} & \multicolumn{3}{c}{Unknown region} \\ \cline{3-8}
 & & Pink & Yellow & Blue & Pink & Yellow & Blue \\ \hline
Input                  & 0.3716 & 0.3223 & 0.1391 & 0.3987 & 0.3508 & 0.1422 & 0.3993 \\
WB (XYZ)           & \bf0.0000 & 0.0382 & 0.1387 & 0.0447 & 0.0287 & 0.1345 & 0.0323 \\
WB (Bradford)    & \bf0.0000 & 0.0274 & 0.1230 & 0.0302 & 0.0134 & 0.1185 & 0.0203 \\
Proposed           & 0.0483 & \bf0.0000 & \bf0.0000 & \bf0.0000 & \bf0.0039 & \bf0.0033 & \bf0.0153 \\
\noalign{\vskip.5mm}
\hline
\end{tabular}
}
\end{center}
\end{table}
\begin{table}[tb]
\caption{CIEDE2000 hue difference $\Delta H$ (post-it note).}
\label{table:2}
\begin{center}
\scalebox{0.75}{
\begin{tabular}{l|c|ccc|ccc}
\hline
\noalign{\vskip.5mm}
\multirow{2}{*}{Method} & \multirow{2}{*}{White} &
\multicolumn{3}{c|}{Known region} & \multicolumn{3}{c}{Unknown region} \\ \cline{3-8}
 &  & Pink & Yellow & Blue & Pink & Yellow & Blue \\ \hline
Input                                      & 9.4615 & 31.7843 & 31.0816 & 33.1196 & 30.7124 & 30.0463 & 30.3340 \\
WB (XYZ)                               & \bf0.0000 & 2.6018 & 6.0621 & 1.9616 & 2.0011 & 6.2878 & 1.1469 \\
WB (Bradford)                        & \bf0.0000 & 1.4322 & 2.2214 & 2.8216 & 0.9348 & 1.5584 & 1.4813 \\
Proposed                                & 5.7288 & \bf0.0000 & \bf0.0000 & \bf0.0000 & \bf0.9105 & \bf0.3368 & \bf1.0764 \\
\noalign{\vskip.5mm}
\hline
\end{tabular}
}
\end{center}
\end{table}
%
%
Figure \ref{fig:result-image} shows the color correction results for ``Post-it note."
Note that we selected pink, yellow, blue post-it notes as target colors and used $(0.7482,0.6855,0.9442)$, $(0.6105,0.7925,0.1208)$, $(0.5832,0.6004,0.9365)$ as the benchmark colors of pink, yellow, blue post-it notes, respectively.
In Table \ref{table:1}, the proposed method was compared with two while-balance adjustments in terms of reproduction angular errors.
From the table, the existing white-balance adjustments did not correct colors except white.
In contrast, the proposed method perfectly corrected the three target-colors in the known target-color region, and reduced the color differences in the unknown target-color region.
From Table \ref{table:2}, the hue difference was also confirmed to have a similar trend to Table \ref{table:1}.
%
\begin{figure*}[b]
\begin{center}
\includegraphics[keepaspectratio, scale=0.4]{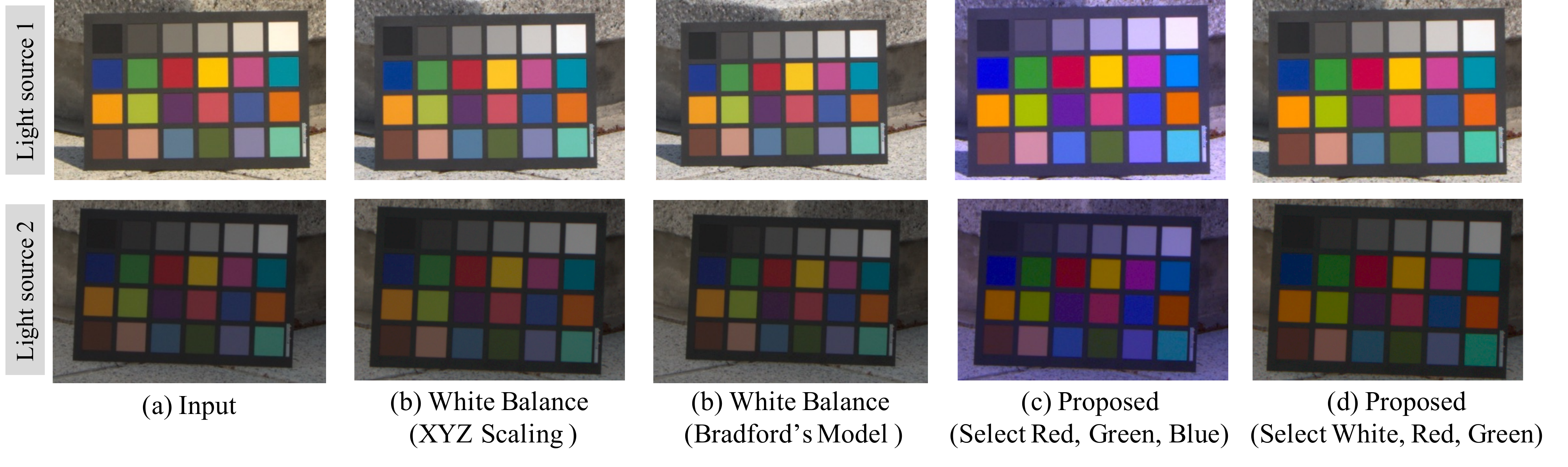}
\end{center}
\caption{Experimental results for ``Color checker."}
\label{fig:exp_results_ColorChecker}
\end{figure*}
%

Figure \ref{fig:exp_results_ColorChecker} shows color correction results for ``Color checker,'' where the images do not have the unknown target-color region.
We prepared two sets of target colors: red, green, blue and white, red, green.
We used $(0.3893,0.1986,0.1124)$, $(0.3454,0.6006,0.1923)$, $(0.1804,0.0722,0.9503)$ as benchmark colors of red, green and blue, respectively.
We also selected the CIE standard illuminant D65 for the benchmark color of white.
As shown in Table \ref{table:3}, the proposed method perfectly adjusted target colors to their benchmark colors; in comparison, the two white balance adjustments corrected only white.
Furthermore, when we select white as one of the target colors, the proposed method has the same correction performance of white as that of white-balancing as confirmed in Table \ref{table:4}.

%
%
\begin{table}[tb]
\caption{Reproduction angular error (color checker).}
\label{table:3}
\begin{center}
\begin{tabular}{l|cccc}
\hline
\noalign{\vskip.5mm}
Method & White & Blue & Green & Red \\ \hline
Input                     & 0.0798 & 0.0149 & 0.0917 & 0.0796 \\
WB (XYZ)              & \bf0.0000  & 0.0679 & 0.0385 & 0.0497 \\
WB (Bradford)       & \bf0.0000 & 0.0503 & 0.0407 & 0.0438 \\
Proposed (R, G, B) & 0.0481 & \bf0.0000 & \bf0.0000 & \bf0.0000 \\
Proposed (W, R, G) & \bf0.0000 & 0.0270 & \bf0.0000 & \bf0.0000 \\
\noalign{\vskip.5mm}
\hline
\end{tabular}%
\end{center}
\end{table}
\begin{table}[tb]
\caption{CIEDE2000 hue difference $\Delta H$ (color checker).}
\label{table:4}
\begin{center}
\begin{tabular}{l|cccc}
\hline
\noalign{\vskip.5mm}
Method & White & Blue & Green & Red \\ \hline
Input                                          & 2.7132 & 0.9592 & 4.9048 & 3.2240 \\
WB (XYZ)                                   & \bf0.0000 & 2.9554 & 1.9783 & 0.6015 \\
WB (Bradford)                            & \bf0.0000 & 0.1868 & 1.4611 & 0.3659 \\
Proposed (R, G, B)                      & 1.0838 & \bf0.0000 & \bf0.0000 & \bf0.0000 \\
Proposed (W, R, G)                      & \bf0.0000 & 3.9246 & \bf0.0000 & \bf0.0000 \\
\noalign{\vskip.5mm}
\hline
\end{tabular}%
\end{center}
\end{table}
%
%
\section{Conclusion}
In this paper, we proposed a multi-color balance adjustment for reducing lighting effects. 
In the proposed method, a color transform matrix is designed by using three target colors and corresponding destination (benchmark) colors.
By applying the matrix to an image, target colors in ``known target-color region'' are perfectly mapped into their benchmark colors.
Additionally, the color differences between benchmark colors and target colors in ``unknown target-color region'' are reduced.

The experimental results showed that the proposed method accurately adjusted three target colors in the known target-color region; in comparison, the conventional white-balance adjustments corrected only white.
The proposed method also reduced the color differences between target colors in unknown target-color regions and corresponding benchmark colors.
%
%
\bibliographystyle{IEEEtran}
\bibliography{references}
 


\end{document}